\documentclass[fleqn,twoside]{article}
\usepackage{espcrc2}

\usepackage{graphicx}
\usepackage[figuresright]{rotating}

\newcommand{\AmS}{{\protect\the\textfont2
  A\kern-.1667em\lower.5ex\hbox{M}\kern-.125emS}}

\title{New methods to measure phase transition strength%
       \thanks{
        WJ and DJ were partially supported by the EC IHP network
        HPRN-CT-1999-000161: {\em Discrete Random Geometries: From Solid
        State Physics to Quantum Gravity\/},
        and
        DJ and RK were partially supported by an
        Enterprise Ireland/British Council Research Visits Scheme.
        RK also wishes
        to thank the TrinLat collaboration for its hospitality during an
        extended stay at Trinity College Dublin.}      
        }

\author{Wolfhard Janke\address[ITP]{Institut f\"ur Theoretische Physik, 
        Universit\"at Leipzig,
        Augustusplatz 10/11, 04109 Leipzig, Germany},
        Des Johnston\address{Department of Mathematics, Heriot-Watt
        University, Riccarton, Edinburgh, EH14 4AS, Scotland}
        and
        Ralph Kenna\address[TCD]{School of Mathematical and Information
        Sciences, Coventry University, Coventry, CV1 5FB, England}
        }

\begin{document}

\begin{abstract}
A recently developed technique to determine the order and strength of 
phase 
transitions by extracting the density of partition function zeroes (a 
continuous function) from finite-size systems (a discrete data set) is 
generalized to systems for which ({\em i\/}) some or all of the zeroes 
occur in 
degenerate sets and/or ({\em ii\/}) they are not confined to a singular 
line in 
the complex plane. The technique is demonstrated by application to the 
case of free Wilson fermions.
\vspace{1pc}
\end{abstract}

\maketitle

\section{INTRODUCTION}
\label{Introduction}

Lattice regularization of a quantum field theory renders the system a 
statistical mechanical one. The issues of phase transitions and their
properties become, therefore, of central importance. While a true phase 
transition can only occur for a system of infinite extent, the 
non-perturbative computational approach to lattice field theory 
and statistical physics accesses only systems of limited size.
The partition function of such a system can be written as 
a polynomial in an appropriate temperature-like or field-like 
variable and the complex zeroes of such a polynomial encode all of the
information on the behaviour of standard thermodynamic quantities.

Traditional statistical mechanical techniques involving
partition function zeroes are mainly confined to analysing the
zeroes closest to the real axis, as these are
the strongest contributors to the critical or pseudocritical
behaviour. However, a full
understanding of the critical properties of the infinite-size system
requires knowledge of the density of zeroes too. It has long been known
that the density of zeroes for a {\emph{finite}}-size system would 
provide 
a lucrative source of information but a reliable technique for the 
extraction of this quantity from numerical data proved elusive. 

Recently, however, some of us have succeeded in providing just such a 
technique \cite{usFisher,usLY}. The general idea is to focus on the 
integrated density of zeroes rather than directly on the density 
itself, 
which, for a finite system, is a series of delta-functions.
This new method has been seen to be quite reliable and robust and is 
applicable to phase transitions of the temperature-driven 
\cite{usFisher}
and field-driven \cite{usLY} types and to transitions of first and 
higher
order. For a comparison between this and  other approaches, see 
\cite{AlHa02}.

For systems hitherto analysed, the zeroes of the partition function 
had two special properties, which seem to be common to
the bulk of standard models encountered in statistical physics.
These are ({\em i\/}) the zeroes are all simple zeroes (zeroes of order 
one)
and ({\em ii\/}) they lie on a curve called the {\emph{singular line}}, 
which
impacts on to the real  axis at the phase transition point.

The question now arises as to the generality of the techniques developed
in \cite{usFisher,usLY}.
Here, we show how the methods can indeed be extended to systems for 
which 
some or all of the zeroes occur in degenerate sets and/or they are 
not confined to a singular line, but instead form a two-dimensional 
pattern in the complex plane.  
Such two-dimensional patterns of zeroes have been observed in 
various lattice field theory and statistical physics models 
\cite{2D84,More2D}. 
Here, we demonstrate the extended technique by 
application to the case of free Wilson fermions.
The zeroes of this lattice field theory 
display the two new features
we wish to address.

The partition function for a lattice of finite extent, $L$, is
$ Z_L(z) \propto  \prod_{j}{\left(z-z_j(L)\right)}$,
where $z$ is an appropriate coupling parameter.
In the case where the zeroes, $z_j$, are 
on a singular line impacting on to the real axis 
at the critical point, $z_c$, they can be
parameterised by
$z_j=z_c+r_j \exp{(i \varphi)}$. The density is then
defined as
$
 g_L(r) = L^{-d} \sum_{j} \delta(r - r_j(L))
$.
The cumulative distribution function of zeroes is
$
 G_L(r)
 =
 \int_0^r{ g_L(s) d s}
$,
which is $j/L^d$ if $ r \in (r_j,r_{j+1})$.
At a zero one assumes the cumulative density is given by the average
\begin{equation}
 G_L(r_j) =  (2j-1)/2L^d \quad .
\label{finite}
\end{equation}
In the thermodynamic limit, for a second-order transition,
the integrated density is, in fact \cite{Abe},
\begin{equation}
 G_\infty(r) \propto  r^{2-\alpha} 
\quad ,
\label{2nd}
\end{equation}
where $\alpha$ is the usual critical exponent associated with specific 
heat.
Standard finite-size scaling emerges quite naturally from this
approach \cite{usFisher}.

\section{GENERAL DISTRIBUTIONS OF ZEROES}
\label{theory}

A departure from smooth linear sets of zeroes was found in 1984 when 
it was shown that for anisotropic two-dimensional lattices there can 
exist 
a two-dimensional distribution (area) of zeroes \cite{2D84}. 
Since then, a host of systems have been discovered with this feature.
A common characteristic of all
two-dimensional distributions of zeroes is that they cross the
 physically 
relevant real axis at only one point, in the thermodynamic limit,
corresponding to the phase transition. 

For such two-dimensional distributions, the density of zeroes 
in the infinite-volume limit has been shown to be \cite{St87}
\begin{equation}
 g(x,y) = y^{1-\alpha-n} f\left( \frac{x}{y^n}\right)
\quad,
\label{2Dinfinite}
\end{equation}
where $(x,y)$ give the location of zeroes with the
critical point as the origin. 
Integrating out the $x$-dependence yields \cite{St87}
\begin{equation}
 g(y) = \int_{x_1}^{x_2}{g(x,y) dx}
 \propto y^{1-\alpha}
\quad ,
\end{equation}
where $x_1$ and $x_2$ mark the extremities of the distribution
of zeroes at a distance $y$ from the $x$ axis.
Integrating again, to determine the cumulative density of zeroes
at the point $r$ in the  $y$-direction, yields
an expression identical to (\ref{2nd}).
Thus, the strength of the transition, as measured by $\alpha$,
can be determined by similar methods to those previously used.
Rather than counting the zeroes along the singular line, one now counts
them up to a line $y=r$ within the two-dimensional domain they inhabit.

The second new feature we wish to accommodate is the existence
of degeneracies in the set of zeroes. If a number of zeroes
coincide, $G_L$, as defined in (\ref{finite}), is multivalued
and is no longer a proper function. 
A more appropriate density function is determined
as follows.
Suppose, in general, that $z_j=z_{j+1}= \dots = z_{j+n-1}$ are $n$-fold
degenerate. 
It is easy to convince oneself that the densities to the left and 
right of
an actual zero are
\begin{equation}
G_L(r) = \left\{
                \begin{array}{ll}
                j+n-1 & \mbox{for $r \in (r_{j+n-1},r_{j+n})$} \\
                j-1   & \mbox{for $r \in (r_{j-1}  ,r_j    ).$}
                \end{array}
         \right.
\label{lr}
\end{equation}
The density at the $n$-fold degenerate zero, $r_j$, is again 
sensibly defined as an average:
\begin{equation}
 G_L(r_j) = \frac{1}{L^d}
                 \left(
                        j+\frac{n}{2}-1
                 \right)   
\label{soln}
\quad .           
\end{equation}
This is the most general formula for extracting the density
of any distribution of zeroes and deals with two-dimensional 
spreads and degeneracies.

\section{APPLICATION}
\label{application}
We wish to test the above technique in a situation with the two new
features of interest -- namely a two-dimensional set of degenerate 
zeroes.
The free Wilson fermion model is an ideal testing ground and its zeroes
in two dimensions are easily generated exactly \cite{KeSe02}.
The zeroes for a system of size $L=40$ are depicted in Fig.~1 in the
complex $z=1/2\kappa$ plane. Here $\kappa$ is the hopping parameter.
\begin{figure}[t]
\vspace{5cm}
\includegraphics{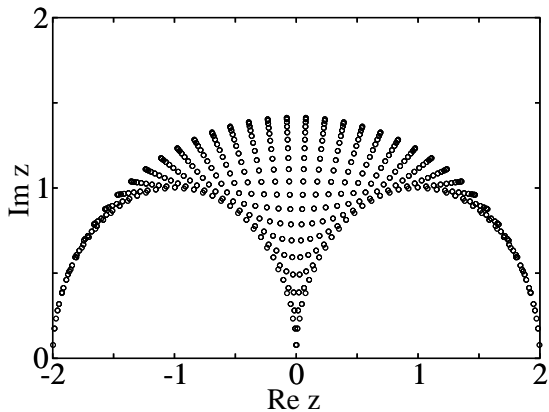}
\caption[a]{The partition function zeroes for free Wilson fermions 
on a $40\times 40$ lattice, where $z=1/2\kappa$ is the inverse hopping
parameter.}
\end{figure}

We generated zeroes for a wide range of lattices, and their 
distributions,
as given by (\ref{soln}), are
plotted in Fig.~2. The data collapses onto a universal curve which
goes through the origin, indicating that (\ref{soln}) is indeed a 
suitable
form for the density of zeroes and 
indicating the occurence of a phase transition.
Fits  close to the origin yield an exponent
compatible with the expected value,  $\alpha = 0$.
The error estimates appropriate to such a fit are non-trivial and we 
leave their discussion to a separate publication \cite{uslater}.

\begin{figure}[t]
\vspace{5cm}
\includegraphics{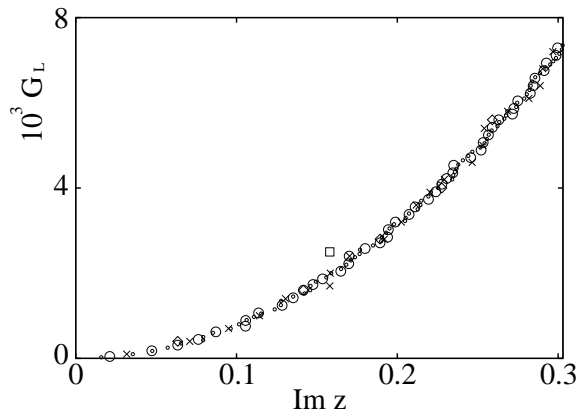}
\caption[a]{
The distribution of zeroes for free Wilson fermions.
The symbols 
$\put(1,0){\framebox(5,5)}~~$, 
{\large{$\diamond$}}, 
$\times$, 
$\put(4,3){\circle{5}}~~$
and
$\put(4,3){\circle{3}}~~$ 
correspond to square lattices of length 
$L=20$, 50, 100, 150, and 200, respectively.
}
\end{figure}

\section{CONCLUSIONS}
\label{conclusions}  

A new method to extract the (continuous) density of zeroes
from (discrete) finite-size data has been extended to deal with the
case of two-dimensional distributions of zeroes and systems in which 
the
zeroes occur in degenerate sets.
The method has been demonstrated in an application to the free Wilson
fermion model and seen to be capable of direct determination of the 
strength
of the phase transition as  measured by the critical exponent 
$\alpha$. 


\end{document}